\documentclass[twocolumn]{article}

\usepackage{comment}
\usepackage{amsmath}
\usepackage{lipsum}
\usepackage{natbib}
\newcommand{\nop}[1]{}

\newcommand*{\ie}{{\em i.e.}}
\usepackage{mathpazo} 
\usepackage[scaled=.80]{helvet}
\usepackage{courier}

\newcommand\hide[1]{}

\usepackage{latexsym}
\usepackage{booktabs}
\usepackage{multirow}
\usepackage{multicol}
\usepackage{hyperref}
\usepackage{url}
\usepackage{enumitem}
\usepackage[ruled,vlined,commentsnumbered]{algorithm2e}
\usepackage{amsfonts}
\usepackage{subcaption}
\usepackage{dblfloatfix}

\usepackage{pgfplots}
\usetikzlibrary{pgfplots.groupplots}
\selectcolormodel{cmyk}

\newenvironment{customlegend}[1][]{%
    \begingroup
    \csname pgfplots@init@cleared@structures\endcsname
    \pgfplotsset{#1}%
}{%
    \csname pgfplots@createlegend\endcsname
    \endgroup
}%

\def\addlegendimage{\csname pgfplots@addlegendimage\endcsname}

\pgfplotsset{
    jitter/.style={
        y filter/.code={\pgfmathparse{\pgfmathresult+rnd*#1}}
    },
    jitter/.default=0.05
}
\pgfplotsset{every tick label/.append style={font=\scriptsize}}
\pgfplotsset{compat=newest}
\pgfdeclarelayer{background}
\pgfdeclarelayer{foreground}
\pgfsetlayers{background,main,foreground}   

\definecolor{TolDarkPurple}{HTML}{332288}
\definecolor{TolDarkBlue}{HTML}{6699CC}
\definecolor{TolLightBlue}{HTML}{88CCEE}
\definecolor{TolLightGreen}{HTML}{44AA99}
\definecolor{TolDarkGreen}{HTML}{117733}
\definecolor{TolDarkBrown}{HTML}{999933}
\definecolor{TolLightBrown}{HTML}{DDCC77}
\definecolor{TolDarkRed}{HTML}{661100}
\definecolor{TolLightRed}{HTML}{CC6677}
\definecolor{TolLightPink}{HTML}{AA4466}
\definecolor{TolDarkPink}{HTML}{882255}
\definecolor{TolLightPurple}{HTML}{AA4499}
\pgfplotscreateplotcyclelist{mbarplot cycle}{%
  {draw=TolDarkBlue,    fill=TolDarkBlue!70},
  {draw=TolLightBrown,  fill=TolLightBrown!70},
  {draw=TolLightGreen,  fill=TolLightGreen!70},
  {draw=TolDarkPink,    fill=TolDarkPink!70},
  {draw=TolDarkPurple,  fill=TolDarkPurple!70},
  {draw=TolDarkRed,     fill=TolDarkRed!70},
  {draw=TolDarkBrown,   fill=TolDarkBrown!70},
  {draw=TolLightRed,    fill=TolLightRed!70},
  {draw=TolLightPink,   fill=TolLightPink!70},
  {draw=TolLightPurple, fill=TolLightPurple!70},
  {draw=TolLightBlue,   fill=TolLightBlue!70},
  {draw=TolDarkGreen,   fill=TolDarkGreen!70},
}
\pgfplotscreateplotcyclelist{mlineplot cycle}{%
  {TolDarkBlue, mark=*, mark size=1.5pt},
  {TolLightBrown, mark=square*, mark size=1.3pt},
  {TolLightGreen, mark=triangle*, mark size=1.5pt},
  {TolDarkBrown, mark=diamond*, mark size=1.5pt},
}
\pgfplotsset{
  mlineplot/.style={
    mbaseplot,
    xmajorgrids=true,
    ymajorgrids=true,
    major grid style={dotted},
    axis x line=bottom,
    axis y line=left,
    legend style={
      cells={anchor=west},
      draw=none
    },
    cycle list name=mlineplot cycle,
  },
  mbarplot base/.style={
    mbaseplot,
    bar width=6pt,
  },
  mbarplot/.style={
    mbarplot base,
    ybar,
    xmajorgrids=false,
    ymajorgrids=true,
    area legend,
    legend image code/.code={%
      \draw[#1] (0cm,-0.1cm) rectangle (0.15cm,0.1cm);
    },
    cycle list name=mbarplot cycle,
  },
  horizontal mbarplot/.style={
    mbarplot base,
    xmajorgrids=true,
    ymajorgrids=false,
    xbar stacked,
    area legend,
    legend image code/.code={%
      \draw[#1] (0cm,-0.1cm) rectangle (0.15cm,0.1cm);
    },
    cycle list name=mbarplot cycle,
  },
  mbaseplot/.style={
    legend style={
      draw=none,
      fill=none,
      cells={anchor=west},
    },
    x tick label style={
      font=\footnotesize
    },
    y tick label style={
      font=\footnotesize
    },
    legend style={
      font=\footnotesize
    },
    major grid style={
      dotted,
    },
  },
  disable thousands separator/.style={
    /pgf/number format/.cd,
      1000 sep={}
  },
}

\usepackage{tikz}
\usetikzlibrary{arrows, decorations.markings, shapes.arrows, patterns, hobby, fit, positioning, shapes, backgrounds, spy}

\usepackage{tikz-qtree}
\usepackage{bm}

\tikzset{textnode/.style={rectangle, inner sep=0pt,outer sep=0,execute at begin node={\strut}, font=\small}}
\tikzset{bnode/.style={circle, draw, fill=black, minimum size=4mm, text=white, outer sep=1.5pt}}
\tikzset{enode/.style={circle, draw, fill=gray!20, minimum size=4mm, inner sep=0.25pt, outer sep=1.5pt}}
\tikzset{inode/.style={circle, thick, draw, minimum size=2mm, outer sep=1.5pt}}
\tikzset{nt/.style={draw, inner xsep=1.5, fill=gray!5, minimum size=3mm, minimum size=1mm, outer sep=1.5pt}}
\tikzset{tnode/.style={minimum size=5mm, font=\large}}
\tikzset{hnode/.style={enode, very thick, draw=blue, outer sep=1.5pt}}
\tikzset{graphletnode/.style={circle, draw, fill=gray!70, minimum size=1.5mm,inner sep=0pt,outer sep=0.25pt}}
\tikzset{hidden/.style={draw=white}}
\tikzset{edge/.style={thick}}
\tikzset{iedge/.style={edge, ultra thick, draw=blue}}
\tikzset{bedge/.style={edge, draw=red}}
\tikzset{faded/.style={opacity=0.60, text opacity=0.60}}
\tikzset{ledge/.style={thick}}

\usepackage{amsmath,amssymb,amsfonts}
\usepackage{algorithmic}
\usepackage{graphicx}
\usepackage{textcomp}
\usepackage{xcolor,colortbl}
\def\BibTeX{{\rm B\kern-.05em{\sc i\kern-.025em b}\kern-.08em
    T\kern-.1667em\lower.7ex\hbox{E}\kern-.125emX}}
    
\usetikzlibrary{calc,shadings,patterns,tikzmark}

\usepackage{graphicx}
\newcommand\update[1]{#1}

\begin{document}
\title{\textbf{Modeling Graphs with Vertex Replacement Grammars}}

\author{Satyaki Sikdar \qquad Justus Hibshman \qquad Tim Weninger \\ 
        Department of Computer Science \& Engineering \\
        University of Notre Dame\\ 
        Notre Dame, IN, USA \\
        \texttt{\{ssikdar,jhibshma,tweninge\}@nd.edu}}

\date{} 

\maketitle

\begin{abstract}
One of the principal goals of graph modeling is to capture the building blocks of network data in order to study various physical and natural phenomena. Recent work at the intersection of formal language theory and graph theory has explored the use of graph grammars for graph modeling. However, existing graph grammar formalisms, like Hyperedge Replacement Grammars, can only operate on small tree-like graphs. The present work relaxes this restriction by revising a different graph grammar formalism called Vertex Replacement Grammars (VRGs). We show that \update{a variant of the} VRG \update{ called Clustering-based Node Replacement Grammar (CNRG)} can be efficiently extracted from many hierarchical clusterings of a graph. We show that \update{CNRGs} encode a succinct model of the graph, yet faithfully preserves the structure of the original graph. In experiments on large real-world datasets, we show that graphs generated from the \update{CNRG} model exhibit a diverse range of properties that are similar to those found in the original networks.
\end{abstract}


\section{Introduction}
We consider the task of identifying the informative and interesting patterns found in graphs. 
Because of their ability to represent natural phenomena, graphs have been studied extensively in various computing and scientific scenarios. 
Arguably the most prescient task in the study of graphs is the identification, extraction, and representation of the small substructures that, in aggregate, describe the underlying phenomenon encoded by the graph. These extracted models contain the LEGO-like building blocks of real-world graphs, and their overarching goal is to enable in-depth scientific analysis and make predictions about the data. 

Because of the prevalence of relevant data and the importance of this line of inquiry, there exists a large body of prior work in graph mining. 
Rooted in data mining and knowledge discovery, subgraph mining methods have been developed to identify frequently occurring subgraphs~\citep{grahne2005fast,jiang2013survey}. 
Unfortunately, these early methods have a so-called ``combinatorial explosion'' problem~\citep{thoma2010discriminative} wherein the search space grows exponentially with the pattern size. 
This causes computational headaches and can also return a massive result set that hinders real-world applicability.
Recent work that heuristically mines graphs for prominent or representative subgraphs have been developed in response, but are still limited by their choice of heuristic~\citep{yan2002gspan,nijssen2005gaston,lin2014large,sun2012efficient}. 
Alternatively, researchers characterize a network by counting small subgraphs called graphlets and therefore forfeit any chance of finding larger, more interesting structures~\citep{prvzulj2007biological,marcus2012rage,ahmed2015efficient}. 

Graph generators, like frequent subgraph mining, also find distinguishing characteristics of networks, but go one step further by generating new graphs that ``look like'' the original graph(s). 
What a graph looks like includes local graph properties like the counts of frequent subgraphs, but can also include global graph properties like the degree distribution, clustering coefficient, diameter, and assortativity metrics among many others. 
Early graph generators had parameters that could be tuned to generate graphs with specific desirable properties. 
Additional work in exponential random graphs~\citep{robins2007introduction}, Kronecker graphs~\citep{leskovec2010kronecker,chakrabarti2004r}, Chung-Lu graphs~\citep{chung2002average}, Stochastic Block Models (SBMs)~\citep{karrer2011stochastic}, and their many derivatives~\citep{pfeiffer2012fast,mussmann2014assortativity,baldesi2018spectral,mussmann2015incorporating,kolda2014scalable} create a model from some example graph in order to generate a new graph that has many of the same properties as the original graph.

These graph models look for small pre-defined patterns or frequently reoccurring patterns, even though interesting and useful information may be hidden in latent and infrequent patterns. 
Principled strategies for extracting these complex patterns are needed to discover the precise mechanisms that govern network structure and growth. 

Recent advances in neural networks have produced graph generators based on recurrent neural networks~\citep{you2018graphrnn}, variational autoencoders~\citep{simonovsky2018graphvae}, and generative adversarial networks~\cite{bojchevski2018netgan} each of which have their advantages and disadvantages, which we explore later. Generally speaking, these neural network models are excellent at generating faithful graphs but struggle to provide a descriptive (\ie, explainable) model from which in-depth scientific or data analysis can be performed.

The present work describes \update{CNRG: a \textbf{C}lustering-based \textbf{N}ode \textbf{R}eplacement \textbf{G}rammar (pronounced: "synergy``) a variant of a vertex replacement grammar (VRG)}, which contains graphical rewriting rules that can match and replace graph fragments similar to how a context-free grammar (CFG) rewrites characters in a string. 
These graph fragments represent a succinct description of the building blocks of the network, and the rewiring rules of the \update{CNRG} describe the instructions about how the graph is pieced together. 

Prior work has investigated the relationship between graph theory and formal language theory by extracting Hyperedge Replacement Grammars (HRGs) from the tree decomposition of a graph~\citep{aguinaga2016growing}. 
The HRG framework can extract patterns from small samples of the graph and can generate networks that have properties that match those of the original graph~\citep{aguinaga2018learning}. 
In their typical use-case, HRGs are used to represent and generate graph patterns through hyperedge rewriting rules, where a nonterminal edge in the graph is matched with a left-hand-side (LHS) rule in the HRG and replaced with its corresponding right-hand-side (RHS). 
The composition of an HRG-rule is entirely dependent on the graph's tree decomposition. 
Unfortunately, finding an optimal tree decomposition is both NP-complete and non-unique. 
Heuristic tree decomposition algorithms exist but still do not scale to even moderately sized graphs. 
Furthermore, non-tree like graphs (\ie, graphs with high treewidth) will produce large, clunky grammar rules that are difficult to interpret.

Like HRGs, VRGs have previously been used to model graph processes and generate graphs. 
Rather than replacing nonterminal (hyper)edges with RHS-subgraphs, a VRG replaces {\em vertices} with RHS-subgraphs. 
VRGs represent an interesting complement to HRGs, but there currently does not exist a means by which to extract a VRG from a graph automatically. 
Instead, graph modelers must craft these grammars by hand, which is a time-consuming process and introduces human bias into the process. 
We desire an automatic, scalable, and interpretable extraction algorithm that compactly models the various structures found in the graph.

The present work describes such an algorithm\footnote{Source code can be found in the \href{https://github.com/satyakisikdar/VRG}{Github repository}.} that automatically extracts a \update{CNRG} from any graph. 
Critically, the extraction algorithm does not require a tree decomposition. 
This permits the extractor to be both scalable and immune to problems arising with non-treelike graphs. 
The output of the \update{CNRG} extractor is a graph model with CFG-like production rules. 
We show that the graph model is able to compress the graph better than state-of-the-art graph summarization models and generate graphs more faithfully than many state-of-the-art graph generation methods.   

\section{Preliminaries} \label{sec:prelim}

We begin with a short introduction to the graph grammar formalism and define important terms that are used throughout the remainder of the present work.

\smallskip
\noindent\textbf{Labeled multigraphs.} \quad
A labeled multigraph is a $4$-tuple $H = \langle V, E, \kappa, L\rangle$ where $V$ is the set of vertices; $E \subseteq V \times V$ is the set of edges; $\kappa : E \mapsto \mathbb{Z}^{+}$ is a function assigning multiplicity to edges; $L$ is the set of labels on nodes and edges. 
By default, each edge has a multiplicity value of 1. Although the \update{CNRG} model can be used for directed graphs, the present work treats all graphs as undirected for clarity of prose and illustration.
We use the terms node and vertex interchangeably in the present work.

\smallskip
\noindent\textbf{\update{Clustering-based Node Replacement Grammars (CNRGs).}}  \quad
A \update{CNRG} is a $4$-tuple $G = \langle \Sigma, \Delta, \mathcal{P}, \mathcal{S} \rangle$ where $\Sigma$ is the alphabet of node labels; $\Delta \subseteq \Sigma$ is the alphabet of terminal node labels; $\mathcal{P}$ is a finite set of productions rules of the form $X \rightarrow (R, f)$, where $X$ is the LHS consisting of a nonterminal node (\ie, $X\in \Sigma \setminus \Delta$) with a size $\omega$, and the tuple $(R, f)$ represent the RHS, where $R$ is a labeled multigraph with terminal and possibly nonterminal nodes, and $f \in \mathbb{Z}^+$ is the frequency of the rule, \ie, the number of times the rule appears in the grammar, and $\mathcal{S}$ is the starting graph which is a non-terminal of size $0$. This formulation is similar to node label controlled (NLC) grammar ~\citep{rozenberg1997handbook}, except that the \update{CNRG} used in the present work does not keep track of specific rewiring conditions. Instead, every internal node in $R$ is labeled by the number of boundary edges to which it was adjacent in the original graph. The sum of the boundary degrees is, therefore, equivalent to $\omega$, which is also equivalent to the label of the LHS.

\begin{figure}[t]
    \centering
    \includegraphics{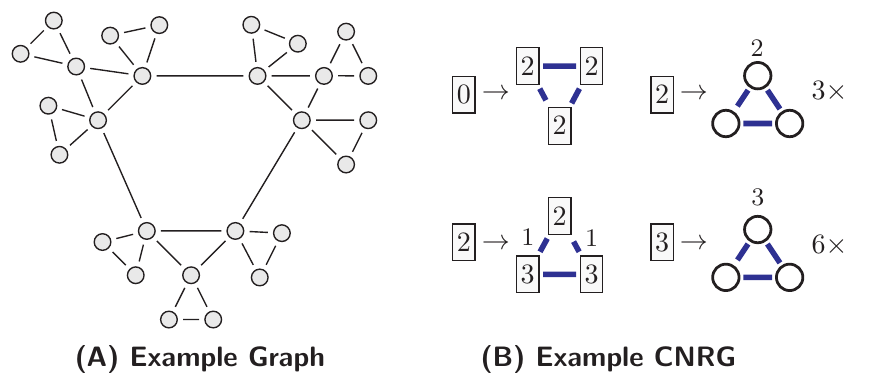}
    \caption{(A) An example graph can be decomposed into a \update{CNRG}. (B) An extracted \update{CNRG} containing four distinct rules, each with an LHS and RHS. The LHS is a single nonterminal node drawn as a square labeled with size $\omega$ (drawn inside the node). The RHS is a subgraph with nonterminal nodes drawn as squares and labeled (illustrated inside the node), terminal nodes labeled with the number of boundary edges (drawn on top of the node), and connecting edges (which do not have labels in this example). The production rules on the right have $f=$ 3$\times$ and $f=$ 6$\times$ indicating that they occur 3 and 6 times respectively.}
    \label{fig:syn}
\end{figure}

A \update{CNRG} can be extracted from any graph or hypergraph and may not be unique. That is, one graph may produce many different \update{CNRG}s. The goal of the present work is to extract \update{CNRG}s that capture the high-order structure of the graph. The example in Fig.~\ref{fig:syn} shows an example graph and an example grammar that can be extracted from it. In this example, the original graph appears to have a regular structure akin to a recursively arranged triangle of triangles. The extracted grammar represents this triangle of triangles pattern, which is represented by the grammar rules.

Like their HRG cousins~\citep{aguinaga2016growing}, the extracted \update{CNRG} may also be used to generate graphs that are similar to (or contain similar high-level structures as) the original graph.

\smallskip
\noindent\textbf{Model size.} \quad One way to compare the conciseness of a grammar is by analyzing its size. For this task, we define a description length (abbreviated as $DL$) for graphs and grammars following prior work. 
Given a labeled multigraph $H$ defined above, we compute its size in the following way. 
Let $\lg (\cdot)$ denote $\log_2(\cdot)$. 
Our approach is similar to that of \citet{cook1993substructure} except that (i) we use Elias $\gamma$~\citep{elias1975universal} encoding instead of the Quinlan \& Rivest encoding~\citep{quinlan1989inferring}, and (ii) we directly encode the multiplicity matrix $M$ instead of encoding a binary adjacency matrix $A$  and its associated multiplicity matrix $M$ separately.
First, $\lg |V|$ and $\lg |L|$ bits are required to encode the number of vertices and the number of labels in $H$ respectively. 
Hence, the total number of bits required to encode all the labeled vertices $(v)$ is $v = \lg |V| + |V| \cdot \lg |L|$ bits. 
Second, let $M$ be a $|V| \times |V|$ multiplicity matrix where $M_{ij} = \kappa(i, j)$ for $(i, j) \in E$, and $0$ otherwise. 
We add $1$ to each element of $M$ to use the $\gamma$-code, which can only encode positive integers. 
Hence, the total number of bits required to encode all the labeled edges $(e)$ is \protect{$e = \lg |E| + \lg |L| \cdot \sum_{ij} |\gamma\text{-code}(M_{ij})|$ bits}. 
Therefore, the description length $(DL(H))$ of the graph $H$ is $DL(H) = (v + e)$ bits. 

Like the graph $H$, the \update{CNRG} $G$ is also given a description length. Each rule ($P$) is of the form $X \rightarrow (R, f)$, where $X$ is a nonterminal of size $\omega$, $R$ is a labeled (multi)graph, and $f$ is the frequency. 
We encode the nonterminal size $\omega$ and the frequency $f$ using the $\gamma$-code. Mathematically, the description length $(DL(l_P))$ of the LHS is given by $DL(l_P) = |\gamma\text{-code}(\omega)| + |\gamma\text{-code}(f)|$ bits.

Similarly, we define a description length $(DL(r_P))$ for the RHS. 
The labeled (multi)graph $R$ is encoded similar to $H$; additionally, we have to include the $\gamma$-encoding of the individual boundary degrees (abbreviated as b\_deg) of the nodes in $V_R$. 
So, we have $DL(r_P) = |\gamma\text{-code}(R)|$ $+ \sum_{v \in V_R} |\gamma\text{-code}(\text{b\_deg}(v))|$ bits.
Therefore, the description length $(DL(G))$ of the \update{CNRG} $G$ is given by $DL(G) = \sum_{P} (DL(l_P) + DL(r_P))$ bits. 

With these definitions formally stated we can more-concretely restate the task: given a (multi)graph $H$, we seek to extract a \update{CNRG} $G$ that succinctly and thoroughly encodes $H$. 
A byproduct of extracting such a graph grammar is that the production rules may also serve as a succinct representation of the constituent structures found in the original graph.

\section{Extracting Vertex Replacement Grammars}

As discussed earlier, many possible CNRGs can represent the same original graph. An optimal CNRG ought to represent the original graph succinctly (\ie, with as few bits as possible) and faithfully (\ie, without losing any information). Unfortunately, such an optimal lossless compression is not possible in all cases. Instead, we assume that $H$ can be clustered hierarchically~\citep{ravasz2003hierarchical} and that regular substructures can be extracted as rules.

The remainder of this section describes the details of several CNRG extraction methods and uses the minimum description length principle to extract a grammar.

\subsection{Hierarchical Graph Clustering}

We begin with a labeled (multi)graph $H$. We first compute a dendrogram from $H$ using a hierarchical clustering algorithm. We explored the Leiden method~\citep{traag2019from}, the Louvain method~\citep{blondel2008fast}, recursive spectral bipartion~\citep{hagen1992new}, and hierarchical spectral $k$-means~\citep{ng2002spectral}; however, any hierarchical clustering method may be used here.

\begin{figure}[t!]
    \centering
    \includegraphics{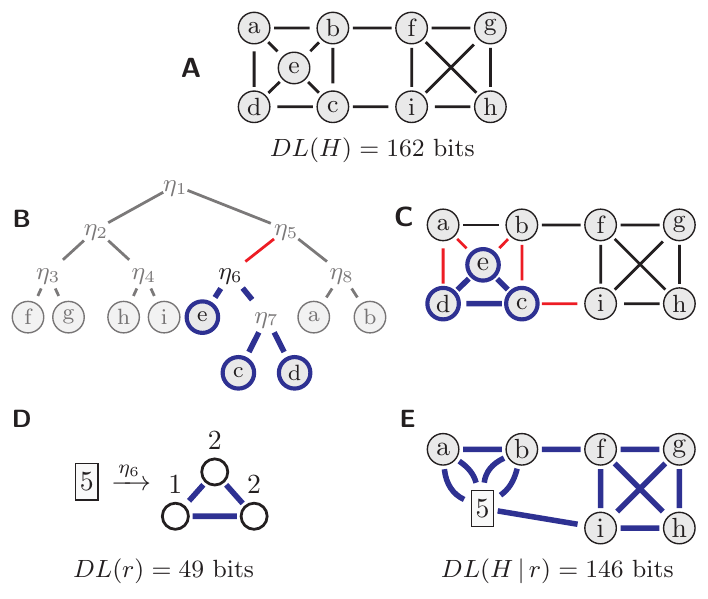}
    \caption{(A) Original graph $H$. (B) Dendrogram created from a hierarchical clustering algorithm, leaves of this dendrogram are nodes of $H$. (C) Subtree $\eta_6$ is selected. Leaf nodes and edges of the induced subgraph are drawn in blue; boundary edges are drawn in red. (D) Rule $\eta_6$ extracted from the $H$. LHS is a nonterminal labeled by $\omega$=5; RHS is the induced subgraph of the nodes in $\eta_6$ labeled with their boundary condition (\ie, number of boundary (red) edges present). (E) New graph $H^\prime$ with $\eta_6$ removed and replaced by a nonterminal node \protect\fbox{5}. }
    \label{fig:ex1}
\end{figure}

As a running example, we introduce a 9-node, 16-edge undirected graph in Fig.~\ref{fig:ex1}(A). 
Applying the recursive spectral clustering algorithm on this graph results in the dendrogram shown in Fig.~\ref{fig:ex1}(B). 
Non-leaf nodes of the dendrogram are represented as $\eta_i$, and the leaves are nodes from the original graph. 
We see that the dendrogram computed from the example graph correctly separates the left and right sides of the graph.
A similar dendrogram is produced when other clustering algorithms are applied. 

\subsection{Rule Extraction}

Given an initial dendrogram $D$ computed by applying a hierarchical clustering algorithm on $H$, the next step is to generate a graph grammar $G$. 
The summary of the rule extraction process is as follows: (a) create production rules from $D$, (b) find the best scoring rule and add it to the grammar $G$, (c) contract the respective subgraphs to create a reduced graph $H^\prime$, and update $D$ to reflect those changes. 
Finally, set $H\gets H^\prime$ and repeat until $D$ is empty.

\smallskip
\noindent\textbf{Creating a Grammar Rule.} \quad Each internal node $\eta \in D$ corresponds to a grammar rule $r_\eta: X \rightarrow (R, f)$. 
Let $V_\eta$ represent the leaf nodes in the subtree rooted at $\eta$, which correspond to nodes in graph $H$. 
Let $b_{\eta}$ represent the set of {\em boundary edges}, \ie, edges in $H$ which have exactly one endpoint in $V_\eta$, and let $\omega = |b_\eta|$. 
$b_\eta$ is used to compute the boundary degrees of the nodes in $V_\eta$.

We set $X$ to be a nonterminal node of size $\omega$ as the LHS of the new production rule. 
The RHS of the new production rule in the CNRG formalism is a labeled multigraph $R\subseteq H$ with rule frequency $f$. 
Let $R = \langle V_R, E_R, \kappa_R, L_R\rangle$ where $V_R = V_\eta$; $E_R = \{(u, v)\ |\ u \in V_\eta \land  v \in V_\eta \land (u, v) \in E \}$; $\kappa_R(e) = k$, where $k$ is the multiplicity of edge $e \in E_R$; $L_R = \{\text{\small{internal node}, \small{internal edge}}\}$. 
If this newly generated rule already exists in the grammar, then the frequency of that rule will be incremented by $1$ (instead of storing duplicate rules). 
Note that this leads to the creation of a many-to-one mapping between the non-leaf nodes of the dendrogram and the rules.
Finally, each subtree, and consequently, each rule, is assigned a score $(s_\eta)$ which is used for selection. The details of the scoring functions are discussed in Sec.~\ref{sec:score}.

\begin{figure}
    \centering
    \includegraphics{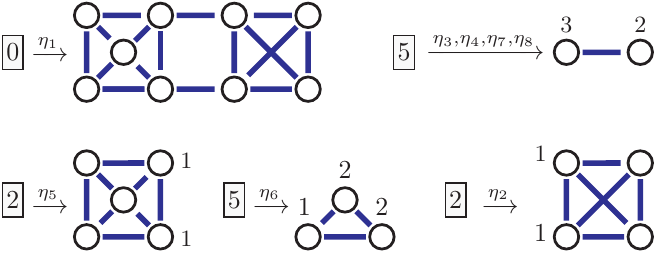}
    \caption{All possible rules that can be extracted from the dendrogram in Fig.~\ref{fig:syn}B labeled by their corresponding subtrees $\eta_{1 \cdots 8}$.}
    \label{fig:dendro_rules}
\end{figure}

Returning to the running example, Fig.~\ref{fig:dendro_rules} shows all possible rules that can be constructed from the dendrogram introduced in Fig.~\ref{fig:ex1}(B). Considering every possible production rule at every step becomes computationally intractable for medium and large-sized graphs. We also observe that certain internal nodes towards the top of the dendrogram cover many leaf nodes and therefore tend to create production rules with large RHSs. Production rules with large RHSs do not align with our aim of finding small, but topologically meaningful building blocks of the graph. So, to prune the search space and restrict the size of the RHS of the rules, we introduce a subtree restriction parameter $\mu$ that removes subtrees larger than $\mu$ from consideration.

\newpage
\noindent\textbf{Selecting the Best Scoring Rule.} \quad From all rules $r_\eta$, we pick the rule $r_\eta^\ast$ with the \emph{minimum} score and add it to the $G$, updating the necessary alphabet $\Sigma$, terminal nodes $\Delta$, and production rules $\mathcal{P}$ as needed. 
Note that multiple subtrees of the dendrogram may correspond to the same rule. For example, in Fig.~\ref{fig:dendro_rules} subtrees $\eta_3, \eta_4, \eta_7$, and $\eta_8$ all correspond to the same rule. 

\smallskip
\noindent\textbf{Updating the Data Structures.}\quad Once a production rule is created from the dendrogram, the next step is to create $H^\prime$ by contracting $H$ by removing the RHS subgraph and inserting the new nonterminal node.

Let $H^\prime = H$ initially. For a selected $\eta^\ast$, we remove $V_\eta$ from $H^\prime$, and insert a new nonterminal node $X$ labeled with $\omega$ (from the first step). We connect $X$ to the rest of the graph through the set of boundary edges in $R$ where edges that were connected to $V_\eta$ are redirected to connect to $X$. Note, this may lead to the creation of multi-edges in the new graph. $H^\prime$ is now strictly smaller than $H$ and contains new nonterminal nodes. 

With a new (smaller) $H^\prime$, it may be prudent to re-run the clustering algorithm and draw a new dendrogram. However, in our initial experiments, we found that re-clustering is time consuming and rarely results in significant changes to the dendrogram. Instead, we simply modify $D$ by replacing the subtrees in $\eta^\ast$  with nonterminal nodes $X$ labeled with  $\omega$. Scores are also updated as needed based on the new graph.

Finally, we set $H\gets H^\prime$ and repeat this process until the dendrogram is empty.

\subsection{Scoring Functions} \label{sec:score}
The choice of scoring function directly impacts the choice of $\eta^*$, which directly impacts the extracted CNRG. Again note that we ignore all $\eta$ where $|V_\eta| > \mu$. The simplest case is to set $s_\eta = |V_\eta| - \mu$. But this simple case results in many ties which need to be broken. For this task we consider three policies:

\begin{itemize}[topsep=0pt, noitemsep]
    \item \textit{Random tiebreaking.} Pick $\eta^\ast$ at random from candidates equi-distant to $\mu$. 
        
    \item \textit{Greedy DL.} Break ties by picking $\eta^\ast$ that minimizes the overall DL of the grammar. Minimizing the DL of the grammar is akin to finding a rule that already exists in the grammar, or by selecting the rule that has the smallest description length among all candidates according to the description length calculation described in Sec.~\ref{sec:prelim}. This is more computationally expensive than other policies because it requires the DL computation for each candidate $\eta$. Among $\eta$'s with equal DL, ties are broken arbitrarily.
        
    \item \textit{Greedy Level.} Break ties by picking $\eta^\ast$ that is at the highest level in the dendrogram. This results in the creation of fewer rules, because a larger portion of the dendrogram, and consequently the graph, is contracted at each step. Among subtrees with equal level, ties are broken arbitrarily.
        
    \item \textit{Greedy level + DL.} Break ties by picking $\eta^\ast$ using the Greedy Level policy first and then by using the DL. 
\end{itemize}

\begin{figure*}[!b]
    \centering
    \setcounter{figure}{4}
    \includegraphics{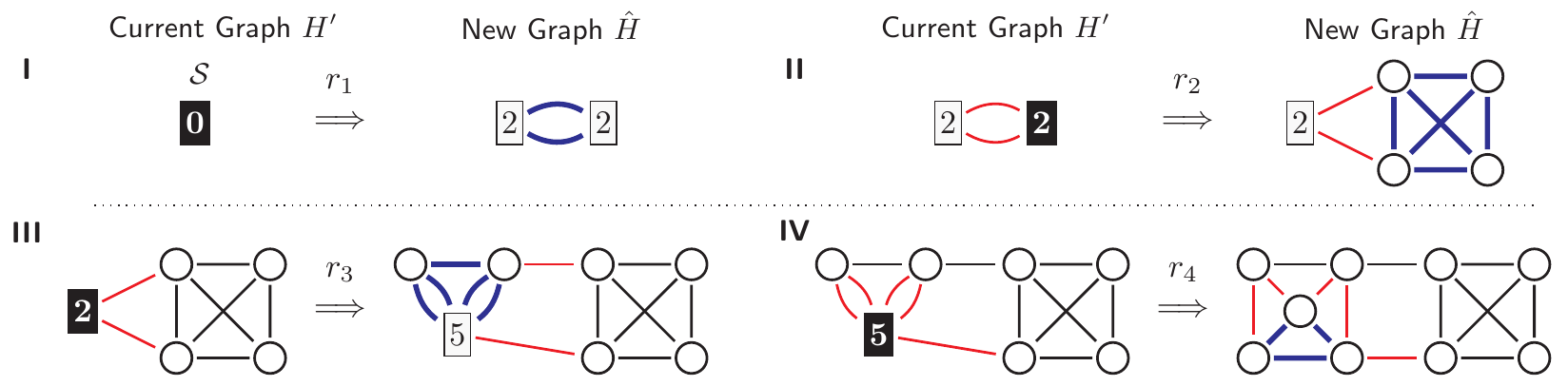}
    \caption{The generation algorithm in action. An application of the rules (in tree-order according to $D$) will regenerate $G$.}
    \label{fig:generation}
\end{figure*}

Previous work suggests that crude two-part MDL~\citep{grunwald2007minimum} is a useful principle for selecting model parameters~\citep{koutra2015summarizing, cook1993substructure}. Therefore, the next policies to select $\eta^\ast$ mimic this. Specifically, let $s_\eta = DL(r_\eta) + DL(H\,|\,r_\eta)$, which is the sum of the DL of the rule and the DL of $H$ compressed by $r_\eta$ respectively. 

Based on this idea, our next task is to calculate $DL(H\,|\,r_\eta)$. One important consideration is the case where multiple subtrees map to the same rule. Again consider the example from Fig.~\ref{fig:dendro_rules} where the subtrees $\eta_3 (\text{f, g}), \eta_4 (\text{h, i}), \eta_7 (\text{c, d}),$ and $\eta_8 (\text{a, b})$ are all encoded in the same rule $r$. With this in mind, two strategies are evident to us: \emph{local MDL} and \emph{global MDL}. In the local MDL strategy, we calculate the scores of each $\eta$ independently, without regard to other subtrees which result in identical rules. In the global strategy, we recognize that identical rules can be compressed together and therefore calculate $DL(H\,|\,r_\eta)$ such that all isomorphic $r_\eta$'s are compressed and stored simultaneously. In global MDL strategy, $\eta^\ast$ is not a single rule, but rather a set of isomorphic rules that are compressed together.

\begin{figure}[t]
    \centering
    \setcounter{figure}{3}
    \includegraphics{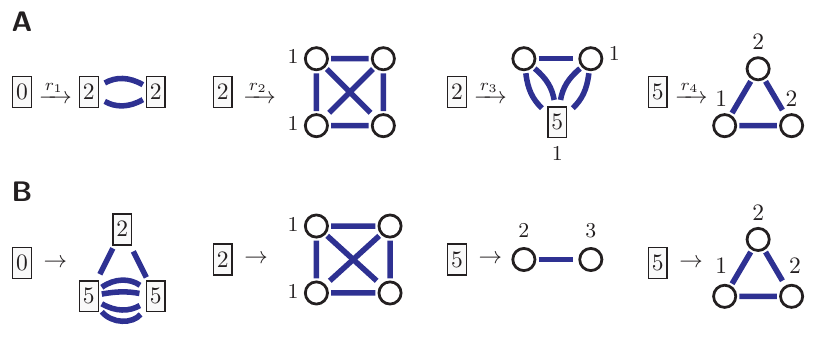}
    \caption{CNRGs obtained from Fig.~\ref{fig:ex1}(B) with $\mu=4$ using (A) the Local MDL strategy and (B) the Global MDL strategy.}
    \label{fig:grammars}
\end{figure}

\setcounter{figure}{5}

We hypothesize that the global MDL strategy will perform best, but requires significantly more time to select $\eta^\ast$. Fig.~\ref{fig:grammars} shows complete CNRGs extracted using the Local (A) and Global (B) MDL strategies. The differences in this small example are subtle. It is unclear which is better.

\section{Generating Graphs from Vertex Replacement Grammars} 

The grammar $G$ encodes information about the original graph $H$ in a way that can be used to generate new graphs. How similar are these newly generated graphs to the original graph? Do they contain similar structures and similar global properties? In this section, we describe how to repeatedly apply rules to generate these graphs. 

We use a stochastic graph generating process to generate graphs. Simply put, this process repeatedly replaces nonterminal nodes with the RHSs of production rules until no nonterminals remain.

Formally, a new graph $H^\prime$ starts with $\mathcal{S}$, a single nonterminal node labeled with $\fbox{0}$. From the current graph, we randomly select a nonterminal and probabilistically (according to each rule's frequency) select a rule from $G$ with an LHS matching the label $\omega$ of the selected nonterminal node. We remove the nonterminal node from $H^\prime$, which breaks exactly $\omega$ edges. Next, we introduce the RHS subgraph to the overall graph randomly rewiring broken edges respecting the boundary degrees of the newly introduced nodes. For example, a node with boundary degree of $3$ expects to be connected with exactly $3$ randomly chosen broken edges. This careful but random rewiring helps preserve topological features of the original network.  After the RHS rule is applied, the new graph $\hat{H}$ will be larger and may have additional nonterminal nodes. We set $H^\prime = \hat{H}$ and repeat this process until no more nonterminals exist.

An example of this generation process is shown in Fig.~\ref{fig:generation} using the rules from Fig.~\ref{fig:grammars}(A). We begin with \fbox{0} and apply $r_1$ to generate a multigraph with two nonterminal nodes and two edges. Next, we (randomly) select the nonterminal on the right and replace it with $r_2$ containing four terminal nodes and 6 new edges. There is one remaining nonterminal, which is replaced with $r_3$ containing two terminal nodes, one nonterminal node, and 5 edges. Finally, the last nonterminal node is replaced with $r_4$ containing three terminal nodes and three edges. The edges are rewired to satisfy the boundary degrees, and we see that $\hat{H}=H$.
In this way, the graph generation algorithm creates new graphs. The previous example conveniently picked rules that would lead to an isomorphic copy of the original graph; however, a stochastic application of rules and random rewiring of broken edges is likely to generate various graph configurations.

\section{Methodology and Results}

Our next task is to evaluate the CNRG model size and its graph generation performance. For size, we measure how the CNRG's description length compares with other graph models. For performance, we measure the accuracy of the stochastic graph generator by comparing the generated graphs with the original graph. 

The goal of the first part of this section is to explore the parameter space for CNRG extraction and generation performance. After we select appropriate parameters, we will compare against existing methods.

\subsection{Datasets}

Datasets were selected based on their variety and size. Our implementation of the CNRG extractor is memory bound at $O(|V|+|E|)$, but it is computationally very fast. The computational complexity of the extractor varies with the choice of clustering algorithm and extractor policy; the graph generation is in $O(|V|+|E|)$. The CNRG extractor can scale to extremely large graphs. Alternative graph models are unable to scale to the largest available graphs, so we selected graphs that could be compared against existing models.
 
We selected five medium-sized graphs from various sources. They are listed in Tab.~\ref{tab:datasets} and were downloaded from KONECT~\citep{kunegis2013konect} and SNAP~\citep{snapnets}.

\begin{table}[t]
    \centering
    \caption{Datasets}
    \begin{tabular}{@{}l rr  r@{}}
        \toprule
        Name & \phantom{a} & $|V|$ & $|E|$  \\
        \midrule
        EuCore Emails && $986$ & $16,687$  \\
        PolBlogs && $1,222$ & $16,717$  \\
        OpenFlights && $2,905$ & $15,645$  \\ 
        ArXiv GrQc && $4,158$ & $13,428$  \\
        Gnutella && $6,299$ & $20,776$  \\
        WikiVote && $7,066$ & $100,736$ \\
        PGP && $10,680$ & $24,316$  \\ 
        \bottomrule
    \end{tabular}
    \label{tab:datasets}
\end{table}

\subsection{Selecting CNRG Parameters}
To measure model size, we must first select from the many parameters of the extraction model: clustering algorithm, boundary information, extractor selection heuristic, and RHS size ($\mu$).

The methodology is as follows. We extract a CNRG for each combination of the clustering algorithm, scoring function, and $\mu \in \{2,3,\ldots,10\}$, which equates to 300 different CNRG models for each dataset. To permit statistical tests and confidence intervals, this process is repeated five times for a total of 1,500 CNRG models for each graph. 

For $k$-way recursive Spectral algorithm, we use $k=\sqrt{n/2}$~\citep{liu2015empirical}. The random hierarchical clustering method split the graph into two (nearly) equally-sized but random clusters in a top-down fashion.

\smallskip
\noindent\textbf{Model Size.}\quad We define the size of the CNRG as the number of rules present in the grammar and its overall complexity. The number of rules is simply the count of the number of distinct production rules. Usually, the grammar size is sufficient to make decisions about the model. Smaller is better.

The description length (DL) measures the size and complexity of the grammar. CNRGs extracted from graphs of different sizes should not be compared in absolute terms -- a large graph will almost certainly have a larger CNRG than a small graph. In order to perform an apples to apples comparison across different dataset sizes, we measure model size using the reciprocal compression ratio: $\textrm{DL}(G)/\textrm{DL}(H)$, where $\textrm{DL}(G)$ is the description length of the CNRG and $\textrm{DL}(H)$ is the description length of the original graph. Lower is better.

\smallskip
\noindent\textbf{Model Performance.}\quad We define the performance of a model as its ability to generate a graph $\hat{H}$ that is similar to the original graph $H$. There are many ways to compare $\hat{H}$ with $H$. In the present work we use the spectral distance ($\lambda$-distance)~\citep{wilson2008study} and \textsc{DeltaCon}~\citep{koutra2016deltacon}.

The $\lambda$-distance compares the spectrum of a graph, which is typically defined as the set of eigenvalues $s = \{\lambda_1, \lambda_2, \ldots, \lambda_{|V|}\}$ are are ordered by their magnitude $\lambda_1 \ge \lambda_2 \ge \ldots \ge \lambda_{|V|}$. The graph spectrum permits a distance to be calculated:

$$\lambda\textrm{-distance}(\hat{H},H) = \sqrt{\sum_{i}{\left(\hat{s_i}-s_i\right)^2}}\textrm{,} $$

\noindent where the list of eigenvalues may be zero-padded if they are not the same size.

\textsc{DeltaCon} measures the difference in node affinities using a belief propagation algorithm. The use of belief propagation implicitly models the diffusion of information throughout the graph and should be able to measure global and local graph structures.

In addition, we count the number of three and four node graphlets~\citep{ahmed2015efficient} that are present in the graph and directly compare these counts. The graphlet correlation distance (GCD) is also used to measure the rank correlation of graphlet orbital counts between nodes in each graph~\citep{prvzulj2007biological,marcus2012rage,hovcevar2014combinatorial}.

Because these are all distance metrics, lower is better.

\smallskip
\noindent\textbf{Selecting a Clustering Method.}\quad First, we consider the selection of a clustering method. We used Random, Leiden, Louvain, recursive spectral bipartition (\ie, Conductance), and hierarchical spectral $k$-means (\ie, Spectral) clustering methods. Each clustering method was applied to each dataset using all available datasets, $\mu$-values and $\eta^\ast$ selection policies. Each unique configuration was repeated 5 times. 

\begin{figure}[tb]
    \centering
    \input{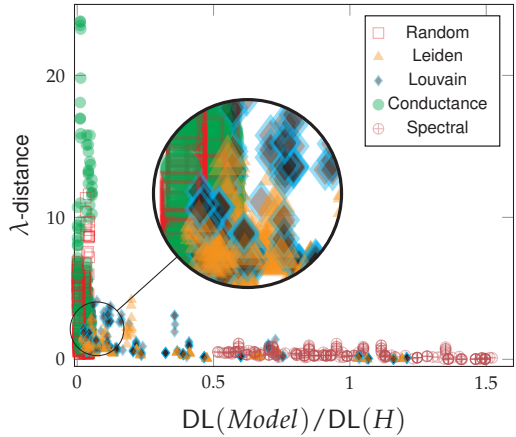}
    \caption{$\lambda$-distance (lower is better) and compression ratio (lower is better) for all runs (all $\mu$, clustering method, $\eta^\ast$ selection policy) on all datasets. Results that are consistently in the bottom-left corner are best. Leiden performs the best consistently. This figure is best viewed in color. }
    \label{fig:clust_sel}
\end{figure}

The model size and $\lambda$-distance for each graph is plotted in Fig.~\ref{fig:clust_sel}.   We observe that Spectral clustering results in remarkably good graph generation, but bad compression. Conversely, Conductance clustering results in remarkably good graph compression, but bad graph generation performance. 

As is typical, we generally observe a trade-off between compression and model performance. The Leiden clustering method appears to perform the best in both metrics consistently; so we select Leiden clustering for further analysis.

\begin{figure*}[t]
    \centering
    \begin{tikzpicture}[font=\sffamily]
\begin{groupplot}[
    group style={
        group name=my plots,
        group size=2 by 1,
        xlabels at=edge bottom,
        xticklabels at=edge bottom,
        vertical sep=15pt,
        horizontal sep=2cm,
    },
    ybar,
    small,
    width=8cm,
    height=4cm,
    /pgf/bar width=3pt,    
    xlabel=\textsf{Clustering Algorithm},
    symbolic x coords={EuCore, Polblogs, Flights, GrQc, PGP},
    xtick=data, 
    xtick style={draw=none},
    y tick label style={
        /pgf/number format/.cd,
            fixed,
            precision=1,
        /tikz/.cd
    },
    ymin=0
]
\nextgroupplot[ylabel={$\textsf{DL}(G) / \textsf{DL}(H)$}, ymax=1]
\addplot+[error bars/.cd,
y dir=both,y explicit] coordinates{
(EuCore, 0.184) +- (EuCore, 0.178)
(Flights, 0.319) +- (Flights, 0.29)
(GrQc, 0.2) +- (GrQc, 0.256)
(PGP, 0.231) +- (PGP, 0.265)
(Polblogs, 0.317) +- (Polblogs, 0.294)
};

\addplot+[error bars/.cd,
y dir=both,y explicit] coordinates{
(EuCore, 0.184) +- (EuCore, 0.178)
(Flights, 0.319) +- (Flights, 0.29)
(GrQc, 0.2) +- (GrQc, 0.256)
(PGP, 0.231) +- (PGP, 0.265)
(Polblogs, 0.317) +- (Polblogs, 0.294)
};

\addplot+[error bars/.cd,
y dir=both,y explicit] coordinates{
(EuCore, 0.184) +- (EuCore, 0.178)
(Flights, 0.319) +- (Flights, 0.29)
(GrQc, 0.2) +- (GrQc, 0.256)
(PGP, 0.231) +- (PGP, 0.265)
(Polblogs, 0.317) +- (Polblogs, 0.294)
};

\addplot+[error bars/.cd,
y dir=both,y explicit] coordinates{
(EuCore, 0.184) +- (EuCore, 0.178)
(Flights, 0.319) +- (Flights, 0.29)
(GrQc, 0.2) +- (GrQc, 0.256)
(PGP, 0.231) +- (PGP, 0.265)
(Polblogs, 0.317) +- (Polblogs, 0.294)
};

\addplot+[error bars/.cd,
y dir=both,y explicit] coordinates{
(EuCore, 0.184) +- (EuCore, 0.178)
(Flights, 0.319) +- (Flights, 0.29)
(GrQc, 0.2) +- (GrQc, 0.256)
(PGP, 0.231) +- (PGP, 0.265)
(Polblogs, 0.317) +- (Polblogs, 0.294)
};

\addplot+[error bars/.cd,
y dir=both,y explicit] coordinates{
(EuCore, 0.184) +- (EuCore, 0.178)
(Flights, 0.319) +- (Flights, 0.29)
(GrQc, 0.2) +- (GrQc, 0.256)
(PGP, 0.59) +- (PGP, 1.061)
(Polblogs, 0.317) +- (Polblogs, 0.294)
};

\nextgroupplot[ylabel={\textsf{$\lambda$-distance}}, ymax=2]
\addplot+[error bars/.cd,
y dir=both,y explicit] coordinates{
(EuCore, 1.535) +- (EuCore, 0.256)
(Flights, 0.731) +- (Flights, 0.155)
(GrQc, 0.523) +- (GrQc, 0.057)
(PGP, 0.845) +- (PGP, 0.135)
(Polblogs, 1.058) +- (Polblogs, 0.244)
};

\addplot+[error bars/.cd,
y dir=both,y explicit] coordinates{
(EuCore, 1.668) +- (EuCore, 0.269)
(Flights, 0.801) +- (Flights, 0.191)
(GrQc, 0.538) +- (GrQc, 0.057)
(PGP, 0.781) +- (PGP, 0.139)
(Polblogs, 1.165) +- (Polblogs, 0.28)
};

\addplot+[error bars/.cd,
y dir=both,y explicit] coordinates{
(EuCore, 1.553) +- (EuCore, 0.261)
(Flights, 0.746) +- (Flights, 0.157)
(GrQc, 0.519) +- (GrQc, 0.055)
(PGP, 0.894) +- (PGP, 0.159)
(Polblogs, 0.858) +- (Polblogs, 0.21)
};

\addplot+[error bars/.cd,
y dir=both,y explicit] coordinates{
(EuCore, 1.613) +- (EuCore, 0.265)
(Flights, 0.666) +- (Flights, 0.12)
(GrQc, 0.508) +- (GrQc, 0.051)
(PGP, 0.813) +- (PGP, 0.149)
(Polblogs, 1.057) +- (Polblogs, 0.262)
};

\addplot+[error bars/.cd,
y dir=both,y explicit] coordinates{
(EuCore, 1.556) +- (EuCore, 0.265)
(Flights, 0.759) +- (Flights, 0.17)
(GrQc, 0.536) +- (GrQc, 0.055)
(PGP, 0.734) +- (PGP, 0.133)
(Polblogs, 0.908) +- (Polblogs, 0.216)
};

\addplot+[error bars/.cd,
y dir=both,y explicit] coordinates{
(EuCore, 1.53) +- (EuCore, 0.259)
(Flights, 0.838) +- (Flights, 0.191)
(GrQc, 0.52) +- (GrQc, 0.051)
(PGP, 0.151) +- (PGP, 0.059)
(Polblogs, 0.873) +- (Polblogs, 0.197)
};

\end{groupplot}
\end{tikzpicture}
    
    \begin{tikzpicture}
    \begin{customlegend}[ 
    legend columns=6,
    legend style={
    draw=none,
    column sep=0.5ex,
    nodes={scale=0.8, transform shape},
  },
  legend entries={\textsf{Random}, \textsf{Greedy Level}, \textsf{Greedy DL}, \textsf{Greedy Level + DL}, \textsf{Local MDL}, \textsf{Global MDL}},
  ]
    \addlegendimage{blue,fill=blue!30!white,mark=square*}
    \addlegendimage{red,fill=red!30!white,mark=square*}
    \addlegendimage{orange,fill=orange!30!white,mark=square*}
    \addlegendimage{black,fill=black!30!white,mark=square*}
    \addlegendimage{violet,fill=violet!90,mark=square*}
    \addlegendimage{green,fill=green!90,mark=square*}
    \end{customlegend}
\end{tikzpicture}
    \caption{Mean model size (left) and graph generation performance (right) for each $\eta^\ast$ selection policy using Leiden clustering. No clear winner is observed. This figure is best viewed in color.}
    \label{fig:type_sel}
\end{figure*}
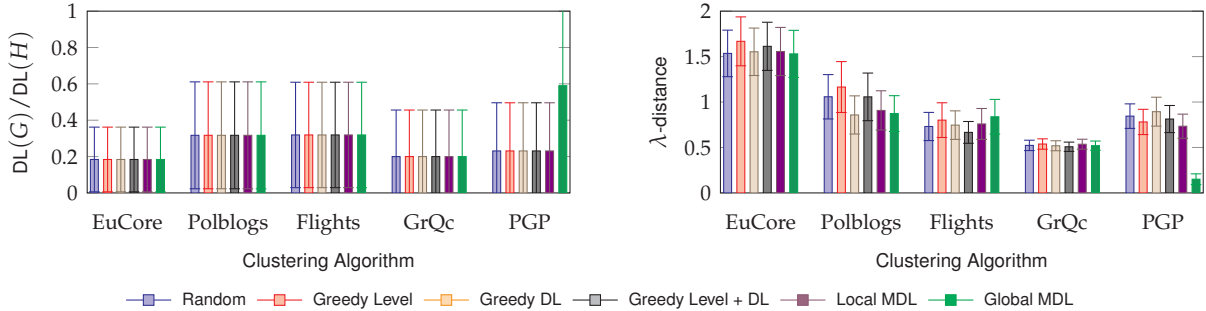
\begin{figure*}[tb]
    \centering
    \begin{tikzpicture}[scale=0.8, transform shape]
\begin{groupplot}[
    group style={
        group name=my plots,
        group size=5 by 2,
        xlabels at=edge bottom,
        xticklabels at=edge bottom,
        vertical sep=10pt,
        horizontal sep=15pt,
    },
    ybar,
    small,
    width=5cm,
    /pgf/bar width=1.75pt,    
    xlabel=$\mu$,
    symbolic x  coords={2, 3, 4, 5, 6, 7, 8, 9, 10},
    xtick=data, 
    xtick style={draw=none},
    y tick label style={
        /pgf/number format/.cd,
            fixed,
            precision=1,
        /tikz/.cd
    },
]

\nextgroupplot[title={\textsf{EuCore}}, ymin=0, ylabel={$\textsf{DL}(Model) / \textsf{DL}(H)$}, ymax=1.2]
\addplot+
coordinates{
(2, 0.735)
(3, 0.366)
(4, 0.182)
(5, 0.103)
(6, 0.057)
(7, 0.055)
(8, 0.052)
(9, 0.052)
(10, 0.052)
};

\addplot+
coordinates{
(2, 0.735)
(3, 0.366)
(4, 0.182)
(5, 0.103)
(6, 0.057)
(7, 0.056)
(8, 0.052)
(9, 0.052)
(10, 0.052)
};

\nextgroupplot[title={\textsf{PolBlogs}}, ymin=0, ymax=1.2]
\addplot+
coordinates{
(2, 1.192)
(3, 0.651)
(4, 0.388)
(5, 0.198)
(6, 0.126)
(7, 0.093)
(8, 0.077)
(9, 0.067)
(10, 0.063)
};

\addplot+
coordinates{
(2, 1.192)
(3, 0.651)
(4, 0.388)
(5, 0.199)
(6, 0.126)
(7, 0.093)
(8, 0.077)
(9, 0.067)
(10, 0.063)
};

\nextgroupplot[title={\textsf{OpenFlights}}, ymin=0, ymax=1.2]
\addplot+
coordinates{
(2, 1.162)
(3, 0.668)
(4, 0.412)
(5, 0.218)
(6, 0.139)
(7, 0.099)
(8, 0.078)
(9, 0.052)
(10, 0.04)
};

\addplot+
coordinates{
(2, 1.162)
(3, 0.668)
(4, 0.412)
(5, 0.218)
(6, 0.139)
(7, 0.099)
(8, 0.078)
(9, 0.052)
(10, 0.04)
};

\nextgroupplot[title={\textsf{GrQc}}, ymin=0, ymax=1.2]
\addplot+
coordinates{
(2, 1.031)
(3, 0.392)
(4, 0.133)
(5, 0.066)
(6, 0.052)
(7, 0.038)
(8, 0.031)
(9, 0.029)
(10, 0.026)
};

\addplot+
coordinates{
(2, 1.031)
(3, 0.392)
(4, 0.133)
(5, 0.066)
(6, 0.052)
(7, 0.038)
(8, 0.032)
(9, 0.029)
(10, 0.026)
};

\nextgroupplot[title={\textsf{PGP}}, ymin=0, ymax=1.2]
\addplot+
coordinates{
(2, 1.063)
(3, 0.475)
(4, 0.232)
(5, 0.108)
(6, 0.07)
(7, 0.044)
(8, 0.035)
(9, 0.027)
(10, 0.022)
};

\addplot+
coordinates{
(2, 1.063)
(3, 0.475)
(4, 0.232)
(5, 0.108)
(6, 0.07)
(7, 0.044)
(8, 0.035)
(9, 0.027)
(10, 0.022)
};

\nextgroupplot[ymin=0, ylabel={\textsf{$\lambda$-distance}}, ymax=5.2]
\addplot+[error bars/.cd,
y dir=both,y explicit] coordinates{
(2, 0.472 ) +- (2, 0.026)
(3, 0.533 ) +- (3, 0.068)
(4, 2.244 ) +- (4, 1.023)
(5, 1.915 ) +- (5, 0.944)
(6, 1.706 ) +- (6, 0.915)
(7, 2.703 ) +- (7, 0.766)
(8, 1.944 ) +- (8, 0.933)
(9, 2.261 ) +- (9, 0.887)
(10, 0.737 ) +- (10, 0.086)
};

\addplot+[error bars/.cd,
y dir=both,y explicit]
coordinates{
(2, 0.475 ) +- (2, 0.018)
(3, 0.485 ) +- (3, 0.048)
(4, 1.59 ) +- (4, 0.909)
(5, 1.987 ) +- (5, 1.018)
(6, 1.453 ) +- (6, 0.86)
(7, 1.997 ) +- (7, 0.955)
(8, 1.766 ) +- (8, 0.992)
(9, 1.713 ) +- (9, 0.93)
(10, 2.536 ) +- (10, 0.782)
};

\nextgroupplot[ymin=0, ymax=5.2]
\addplot+[error bars/.cd,
y dir=both,y explicit]
coordinates{
(2, 0.034 ) +- (2, 0.004)
(3, 0.229 ) +- (3, 0.034)
(4, 0.353 ) +- (4, 0.071)
(5, 3.791 ) +- (5, 1.252)
(6, 0.485 ) +- (6, 0.052)
(7, 0.964 ) +- (7, 0.382)
(8, 1.114 ) +- (8, 0.394)
(9, 1.269 ) +- (9, 0.378)
(10, 1.272 ) +- (10, 0.336)
};

\addplot+[error bars/.cd,
y dir=both,y explicit]
coordinates{
(2, 0.03 ) +- (2, 0.003)
(3, 0.234 ) +- (3, 0.037)
(4, 0.326 ) +- (4, 0.047)
(5, 2.225 ) +- (5, 1.635)
(6, 0.594 ) +- (6, 0.137)
(7, 0.865 ) +- (7, 0.348)
(8, 1.147 ) +- (8, 0.309)
(9, 1.259 ) +- (9, 0.373)
(10, 1.496 ) +- (10, 0.329)
};

\nextgroupplot[ymin=0, ymax=5.2]
\addplot+[error bars/.cd,
y dir=both,y explicit]
coordinates{
(2, 0.066 ) +- (2, 0.024)
(3, 0.235 ) +- (3, 0.072)
(4, 0.456 ) +- (4, 0.061)
(5, 0.585 ) +- (5, 0.095)
(6, 0.74 ) +- (6, 0.101)
(7, 0.738 ) +- (7, 0.063)
(8, 0.805 ) +- (8, 0.028)
(9, 1.501 ) +- (9, 0.927)
(10, 0.866 ) +- (10, 0.075)
};

\addplot+[error bars/.cd,
y dir=both,y explicit]
coordinates{
(2, 0.061 ) +- (2, 0.023)
(3, 0.217 ) +- (3, 0.072)
(4, 0.447 ) +- (4, 0.072)
(5, 0.637 ) +- (5, 0.091)
(6, 0.705 ) +- (6, 0.044)
(7, 0.752 ) +- (7, 0.057)
(8, 0.831 ) +- (8, 0.058)
(9, 2.365 ) +- (9, 1.155)
(10, 0.813 ) +- (10, 0.062)
};

\nextgroupplot[ymin=0, ymax=5.2]
\addplot+[error bars/.cd,
y dir=both,y explicit]
coordinates{
(2, 0.046 ) +- (2, 0.003)
(3, 0.26 ) +- (3, 0.022)
(4, 0.486 ) +- (4, 0.086)
(5, 0.422 ) +- (5, 0.065)
(6, 0.518 ) +- (6, 0.057)
(7, 0.716 ) +- (7, 0.051)
(8, 0.696 ) +- (8, 0.1)
(9, 0.65 ) +- (9, 0.069)
(10, 0.773 ) +- (10, 0.111)
};

\addplot+[error bars/.cd,
y dir=both,y explicit]
coordinates{
(2, 0.044 ) +- (2, 0.006)
(3, 0.26 ) +- (3, 0.021)
(4, 0.455 ) +- (4, 0.068)
(5, 0.56 ) +- (5, 0.093)
(6, 0.534 ) +- (6, 0.108)
(7, 0.752 ) +- (7, 0.038)
(8, 0.701 ) +- (8, 0.054)
(9, 0.702 ) +- (9, 0.071)
(10, 0.812 ) +- (10, 0.153)
};

\nextgroupplot[ymin=0, ymax=5.2]
\addplot+[error bars/.cd,
y dir=both,y explicit]
coordinates{
(2, 0.028 ) +- (2, 0.006)
(3, 0.071 ) +- (3, 0.007)
(4, 0.279 ) +- (4, 0.038)
(5, 0.859 ) +- (5, 0.33)
(6, 1.073 ) +- (6, 0.234)
(7, 0.899 ) +- (7, 0.331)
(8, 1.018 ) +- (8, 0.338)
(9, 1.638 ) +- (9, 0.661)
(10, 1.448 ) +- (10, 0.491)
};

\addplot+[error bars/.cd,
y dir=both,y explicit]
coordinates{
(2, 0.023 ) +- (2, 0.006)
(3, 0.057 ) +- (3, 0.006)
(4, 0.355 ) +- (4, 0.057)
(5, 0.873 ) +- (5, 0.413)
(6, 0.926 ) +- (6, 0.31)
(7, 0.778 ) +- (7, 0.278)
(8, 0.825 ) +- (8, 0.333)
(9, 1.179 ) +- (9, 0.391)
(10, 1.591 ) +- (10, 0.479)
};

\end{groupplot}
\end{tikzpicture}
    \begin{tikzpicture}
    \begin{customlegend}[ 
    legend columns=2,
    legend style={
    draw=none,
    column sep=2ex,
    nodes={scale=0.8, transform shape},
  },
  legend entries={\textsf{Greedy Level + DL}, \textsf{Local MDL}},
  ]
    \addlegendimage{blue,fill=blue!30!white,mark=square*}
    \addlegendimage{red,fill=red!30!white,mark=square*}
    \end{customlegend}
\end{tikzpicture}
    \caption{Mean model size (top), and graph generation performance (bottom) for each $\mu$ using Leiden clustering. We select $\mu$=4 as having the best size-to-performance tradeoff. This figure is best viewed in color.}
    \label{fig:lambda_sel}
\end{figure*}
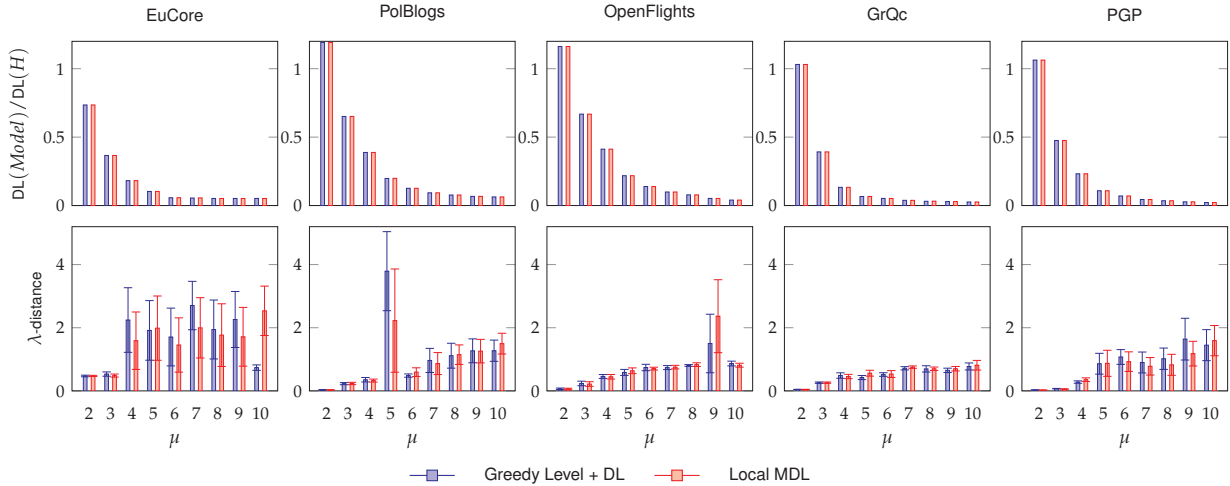

\smallskip
\noindent\textbf{Selecting an $\eta^\ast$ policy.}\quad Our next task is to find the $\eta^\ast$ selection policy that performs best. Using only the Leiden clustering method, we group all runs (across all $\mu$ values) and plot the mean reciprocal compression ratio and $\lambda$-distance in Fig.~\ref{fig:type_sel}. 95\% confidence intervals are drawn as error bars.

We observe that the choice of $\eta^\ast$ selection policy has minimal effect on the model size and the generation performance. We select Greedy Level + DL and Local MDL because they performed (slightly) better than the other methods, but also because they are much faster to compute than the Global MDL.

\smallskip
\noindent\textbf{Selecting a $\mu$ Value.}\quad Next, we compare model size and generation performance results for various values of $\mu$. Recall that $\mu$ is an upper bound for the number of nodes that appear within a subtree of $\eta^\ast$; \ie, $\mu$ is the maximum number of nodes that can appear in any extracted RHS.

We select $\mu$ by following the pattern as before. Using the Leiden clustering method and Greedy Level + DL and Local MDL $\eta^\ast$ selection policies, we plot the mean reciprocal compression ratio and $\lambda$-distance in Fig.~\ref{fig:lambda_sel}. 95\% confidence intervals are drawn as error bars.

We observe little difference between the $\eta^\ast$ selection policies. However, the size-to-performance trade-off becomes evident again as $\mu$ varies from small to large. Small values of $\mu$ more complex models but more accurate models, while larger values produce less complex models but less accurate models; however, there are quickly diminishing returns as $\mu$ increases. 

We select a $\mu$=4 because it appears to generate reasonably small models with reasonable accuracy.

In summary, based on the decisions highlighted in this section we select a parameterization for the CNRG that uses Leiden clustering, the Greedy Level + DL for the $\eta^\ast$ selection policy, and $\mu=4$. We will use these values throughout the remainder of the present work unless otherwise specified.

\begin{table}[t]
    \centering
    \caption{Model Size Comparison. Lower is better.}
    \begin{tabular}{@{}l rrrr@{}}
        \toprule
        \multirow{2}{*}{Graph} & \multicolumn{4}{c}{$DL(Model) / DL(H)$}  \\
        \cmidrule{2-5}
         & SUBDUE & SlashBurn & VoG & CNRG \\
        \midrule
        Karate & $3.546$ & $1.119$ & $1.080$ & $\mathbf{0.704}$ \\
        Dolphins & $4.348$ & $1.336$ & $1.026$ & $\mathbf{0.43}$ \\
        LesMis & $3.546$ & $1.05$ & $\mathbf{0.875}$ & $0.924$ \\
        EuCore & -- & $5.54$ & $0.986$ & $\mathbf{0.182}$ \\
        PolBlogs & -- & $0.873$ & $0.881$ & $\mathbf{0.388}$ \\
        OpenFlights & -- & $0.888$ & $0.869$ & $\mathbf{0.412}$ \\
        GrQc & -- & $1.154$ & $0.851$ & $\mathbf{0.133}$ \\
        PGP & -- & $1.196$ & $0.911$ & $\mathbf{0.232}$ \\
        Gnutella & -- & $1.045$ & $0.967$ & $\mathbf{0.306}$ \\
        WikiVote & -- & $0.839$ & $0.843$ & $\mathbf{0.525}$ \\
        \bottomrule
    \end{tabular}
    \label{tab:model_size}
\end{table}

\subsection{Graph Model Size}

Next, we compare the model size of \update{CNRG}, parameterized as above, in bits against \update{three} other graph models: the Vocabulary-based summarization of Graphs (VoG)~\citep{koutra2015summarizing}, SlashBurn~\citep{lim2014slashburn}, and SUBDUE~\citep{ketkar2005subdue}. Like the \update{CNRG} model, VoG, SlashBurn, and SUBDUE maintain an encoding of the graph, but their models are constructed in very different ways. VoG summarizes graphs using a fixed vocabulary of structures. SlashBurn  recursively splits a graph into hubs and spokes connected only by the hubs. SUBDUE creates a node-grammar model, similar in principle to the \update{CNRG} model, by finding substructures that maximally reduce the size (bits) of the graph after each selection. These models are useful for graph summarizing and graph understanding, but do not generate graphs; thus, they can only be compared to \update{CNRG}s by their model size. Each of the models was run with their default settings. 

SUBDUE was unable to process the even the smallest of our graph datasets, so we included three graphs: Karate, Dolphins, and LesMis, representing well known small graphs, in Tab.~\ref{tab:model_size}. These results show that \update{CNRG} almost always produces the best model sizes among the other models. This confirms our hypothesis that CNRG compresses the original graph better than the state-of-the-art methods.

\subsection{Graph Generation Performance}

Here we show that the CNRG model represents not only a succinct encoding of the original graph but also a faithful one as well. Keeping a tree ordering over production rules in the CNRG will permit a generation close or isomorphic to the original graph. This is an interesting, but not particularly useful outcome of the CNRG model. Instead, we ask how well the CNRG model generates new graphs. Are these graphs similar to the original graph? How does the CNRG accuracy compare to other graph models at generating graphs?

Graph generators have been studied intently for several years. The idea being that we only truly understand a graph if we can generate it faithfully. Practically speaking, graph generators are often used to create null models for statistical purposes. In a similar vein, graph generators are frequently used to find anomalous patterns in real-world graphs. 

\smallskip
\noindent\textbf{Setup.}\quad We compare CNRG graph generation against many of the state-of-the-art graph generators. We consider the properties that characterize some real-world networks and compare the distribution of graphs generated using the Kronecker graph model~\citep{leskovec2010kronecker}, the Block Two-Level Erd\H{o}s-R\'{e}nyi (BTER) model~\citep{kolda2014scalable}, Chung-Lu's configuration model~\citep{chung2002average}, the degree corrected Stochastic Block Model (DC-SBM)~\citep{karrer2011stochastic}, and the stochastic Hyperedge Replacement Grammar (HRG) model~\citep{aguinaga2016growing,aguinaga2018learning}. 

Like CNRGs, these other graph models learn parameters that can be used to approximately recreate the original graph or a graph of some other size such that the generated graph holds many of the same properties as the original graph. The generated graphs are likely not isomorphic to the original graph. We can, however, still judge how closely the generated graph resembles the original graph by comparing several of their local and global graph properties. 

\begin{table*}[htb]
    \centering
        \caption{Graph generation performance. Graphs generated by CNRG closely match the original graph and are consistently the best or close to the best performing model. Lower is better; best results are indicated by boldface.}
    \footnotesize{
    \begin{tabular}{@{}l  rrr | rrr | rrr@{}}
\toprule
& \multicolumn{3}{c}{\textbf{EuCore}} &  \multicolumn{3}{c}{\textbf{PolBlogs}} &  \multicolumn{3}{c}{\textbf{OpenFlights}} \\
    & GCD &  $\lambda$-dist & \sc{DeltaCon} & GCD &  $\lambda$-dist & \sc{DeltaCon} & GCD &  $\lambda$-dist & \sc{DeltaCon}  \\
\midrule
ChungLu & 0.409 &  \textbf{0.803} &    6661 & 0.466 &  \textbf{1.234} &    8020  & 1.1116 &  \textbf{0.614} &    14142 \\
    HRG &  0.229 &  8.091 &    7841 & 1.196 &   4.407 &    8872 & 1.2442 & 2.761 &    15860\\
   DC-SBM & \textbf{0.180} &  2.057 & \textbf{5736}  & 0.262 &  4.186 &    8023& 0.8414 &  3.534 &    11450\\
   BTER & - &  - &    - & 0.352 &   7.505 &    8444&  0.832 &   4.936 &    13269 \\     
   Kronecker & 0.3164 & 11.802 & 4840 & 1.302 & 14.31 & 6140 & 1.83  &  10.459 & \textbf{8589}\\
   VRG & 0.233 &  4.969 &    5793 & \textbf{0.212} &  {4.276} &    \textbf{7436}& \textbf{0.2832} &  3.581 &    11473\\

\rule{0pt}{4ex} 
& \multicolumn{3}{c}{\textbf{GrQc}} &  \multicolumn{3}{c}{\textbf{PGP}} &  \multicolumn{3}{c}{\textbf{Gnutella}} \\ 
    & GCD &  $\lambda$-dist & \sc{DeltaCon} & GCD &  $\lambda$-dist & \sc{DeltaCon} & GCD &  $\lambda$-dist & \sc{DeltaCon}  \\ \midrule
 ChungLu & 2.657 &  \textbf{0.389} &  21607& 2 & \textbf{0.64} & 18503& 1.02 & 0.42  & 34451  \\
      HRG & 1.99 &  4.41 & \textbf{12153}& - & - & - & 2 &  5 & \textbf{20755}\\
      DC-SBM & 2.065 &  2.202 &    14456 & 1.39 & 2.29  & 15216& - & - & - \\
      BTER &  2.231 &   0.439 &    14066 & 1.61 & 0.832  & 15161&  1.10 & 0.474  &  32692\\
      Kronecker & 3.87 & 5.468 & 13173   & 2.882  &  3.54  & 12320& 3.31  & 5.96 & 22145 \\
      VRG &  \textbf{1.067} &   0.723 &    13528 & \textbf{0.448} & 1.329 & \textbf{12257}& \textbf{0.41} & \textbf{0.20} & 30616 \\

\bottomrule
\end{tabular}

    }
    \label{tab:results}
\end{table*}   

\begin{figure}[htb]
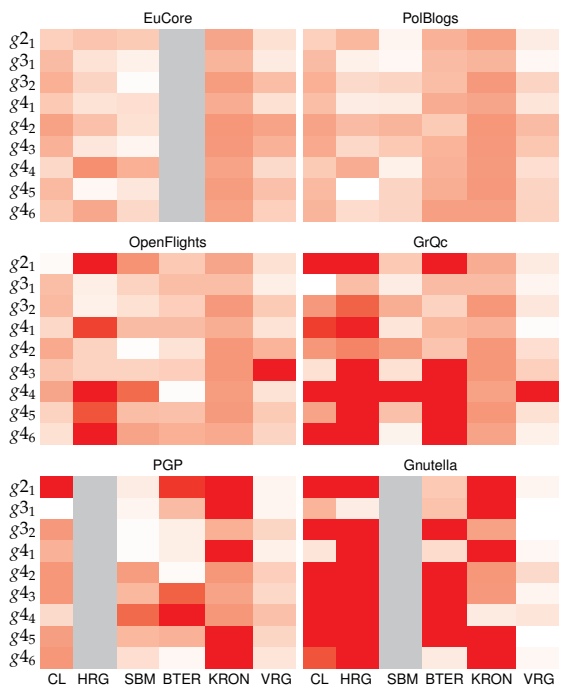

    \centering
    \footnotesize{
    \scalebox{0.85}{
\setlength{\tabcolsep}{0.2em}
\centering
\begin{tabular}{lccccccccccccc}
\rule{0pt}{3ex} & \multicolumn{6}{c}{\textsf{EuCore}} &  & \multicolumn{6}{c}{\textsf{PolBlogs}} \\
$g2_1$ & \cellcolor{red!21}~ & \cellcolor{red!27}~ & \cellcolor{red!24}~ & \cellcolor{black!25}~ & \cellcolor{red!42}~ & \cellcolor{red!13}~ &  & \cellcolor{red!21}~ & \cellcolor{red!33}~ & \cellcolor{red!4}~ & \cellcolor{red!36}~ & \cellcolor{red!44}~ & \cellcolor{red!8}~ \\
$g3_1$ & \cellcolor{red!31}~ & \cellcolor{red!12}~ & \cellcolor{red!6}~ & \cellcolor{black!25}~ & \cellcolor{red!34}~ & \cellcolor{red!9}~ &  & \cellcolor{red!31}~ & \cellcolor{red!6}~ & \cellcolor{red!3}~ & \cellcolor{red!31}~ & \cellcolor{red!34}~ & \cellcolor{red!3}~ \\
$g3_2$ & \cellcolor{red!37}~ & \cellcolor{red!19}~ & \cellcolor{red!1}~ & \cellcolor{black!25}~ & \cellcolor{red!47}~ & \cellcolor{red!30}~ &  & \cellcolor{red!37}~ & \cellcolor{red!16}~ & \cellcolor{red!19}~ & \cellcolor{red!30}~ & \cellcolor{red!49}~ & \cellcolor{red!19}~ \\
$g4_1$ & \cellcolor{red!25}~ & \cellcolor{red!12}~ & \cellcolor{red!14}~ & \cellcolor{black!25}~ & \cellcolor{red!39}~ & \cellcolor{red!13}~ &  & \cellcolor{red!30}~ & \cellcolor{red!8}~ & \cellcolor{red!9}~ & \cellcolor{red!39}~ & \cellcolor{red!42}~ & \cellcolor{red!11}~ \\
$g4_2$ & \cellcolor{red!44}~ & \cellcolor{red!29}~ & \cellcolor{red!13}~ & \cellcolor{black!25}~ & \cellcolor{red!50}~ & \cellcolor{red!43}~ &  & \cellcolor{red!43}~ & \cellcolor{red!31}~ & \cellcolor{red!34}~ & \cellcolor{red!24}~ & \cellcolor{red!50}~ & \cellcolor{red!31}~ \\
$g4_3$ & \cellcolor{red!36}~ & \cellcolor{red!10}~ & \cellcolor{red!4}~ & \cellcolor{black!25}~ & \cellcolor{red!49}~ & \cellcolor{red!36}~ &  & \cellcolor{red!40}~ & \cellcolor{red!17}~ & \cellcolor{red!25}~ & \cellcolor{red!35}~ & \cellcolor{red!49}~ & \cellcolor{red!26}~ \\
$g4_4$ & \cellcolor{red!17}~ & \cellcolor{red!54}~ & \cellcolor{red!37}~ & \cellcolor{black!25}~ & \cellcolor{red!43}~ & \cellcolor{red!16}~ &  & \cellcolor{red!24}~ & \cellcolor{red!39}~ & \cellcolor{red!6}~ & \cellcolor{red!36}~ & \cellcolor{red!48}~ & \cellcolor{red!14}~ \\
$g4_5$ & \cellcolor{red!32}~ & \cellcolor{red!3}~ & \cellcolor{red!10}~ & \cellcolor{black!25}~ & \cellcolor{red!47}~ & \cellcolor{red!29}~ &  & \cellcolor{red!32}~ & \cellcolor{white}~ & \cellcolor{red!19}~ & \cellcolor{red!38}~ & \cellcolor{red!49}~ & \cellcolor{red!19}~ \\
$g4_6$ & \cellcolor{red!26}~ & \cellcolor{red!41}~ & \cellcolor{red!17}~ & \cellcolor{black!25}~ & \cellcolor{red!44}~ & \cellcolor{red!21}~ &  & \cellcolor{red!34}~ & \cellcolor{red!15}~ & \cellcolor{red!19}~ & \cellcolor{red!45}~ & \cellcolor{red!45}~ & \cellcolor{red!20}~ \\
\rule{0pt}{3ex} & \multicolumn{6}{c}{\textsf{OpenFlights}} &  & \multicolumn{6}{c}{\textsf{GrQc}} \\
$g2_1$ & \cellcolor{red!2}~ & \cellcolor{red!100}~ & \cellcolor{red!52}~ & \cellcolor{red!25}~ & \cellcolor{red!42}~ & \cellcolor{red!13}~ &  & \cellcolor{red!100}~ & \cellcolor{red!100}~ & \cellcolor{red!25}~ & \cellcolor{red!100}~ & \cellcolor{red!39}~ & \cellcolor{red!9}~ \\
$g3_1$ & \cellcolor{red!30}~ & \cellcolor{red!7}~ & \cellcolor{red!20}~ & \cellcolor{red!30}~ & \cellcolor{red!30}~ & \cellcolor{red!7}~ &  & \cellcolor{red!0}~ & \cellcolor{red!30}~ & \cellcolor{red!8}~ & \cellcolor{red!31}~ & \cellcolor{red!31}~ & \cellcolor{red!4}~ \\
$g3_2$ & \cellcolor{red!32}~ & \cellcolor{red!6}~ & \cellcolor{red!13}~ & \cellcolor{red!22}~ & \cellcolor{red!49}~ & \cellcolor{red!24}~ &  & \cellcolor{red!49}~ & \cellcolor{red!76}~ & \cellcolor{red!38}~ & \cellcolor{red!20}~ & \cellcolor{red!50}~ & \cellcolor{red!10}~ \\
$g4_1$ & \cellcolor{red!17}~ & \cellcolor{red!89}~ & \cellcolor{red!31}~ & \cellcolor{red!31}~ & \cellcolor{red!38}~ & \cellcolor{red!12}~ &  & \cellcolor{red!90}~ & \cellcolor{red!98}~ & \cellcolor{red!11}~ & \cellcolor{red!33}~ & \cellcolor{red!36}~ & \cellcolor{red!1}~ \\
$g4_2$ & \cellcolor{red!40}~ & \cellcolor{red!20}~ & \cellcolor{red!0.7}~ & \cellcolor{red!12}~ & \cellcolor{red!50}~ & \cellcolor{red!35}~ &  & \cellcolor{red!50}~ & \cellcolor{red!60}~ & \cellcolor{red!47}~ & \cellcolor{red!27}~ & \cellcolor{red!50}~ & \cellcolor{red!14}~ \\
$g4_3$ & \cellcolor{red!27}~ & \cellcolor{red!20}~ & \cellcolor{red!20}~ & \cellcolor{red!22}~ & \cellcolor{red!50}~ & \cellcolor{red!100}~ &  & \cellcolor{red!13}~ & \cellcolor{red!100}~ & \cellcolor{red!13}~ & \cellcolor{red!100}~ & \cellcolor{red!50}~ & \cellcolor{red!21}~ \\
$g4_4$ & \cellcolor{red!42}~ & \cellcolor{red!100}~ & \cellcolor{red!72}~ & \cellcolor{red!1}~ & \cellcolor{red!47}~ & \cellcolor{red!12}~ &  & \cellcolor{red!100}~ & \cellcolor{red!100}~ & \cellcolor{red!100}~ & \cellcolor{red!100}~ & \cellcolor{red!44}~ & \cellcolor{red!100}~ \\
$g4_5$ & \cellcolor{red!20}~ & \cellcolor{red!82}~ & \cellcolor{red!30}~ & \cellcolor{red!29}~ & \cellcolor{red!48}~ & \cellcolor{red!24}~ &  & \cellcolor{red!43}~ & \cellcolor{red!100}~ & \cellcolor{red!29}~ & \cellcolor{red!100}~ & \cellcolor{red!50}~ & \cellcolor{red!13}~ \\
$g4_6$ & \cellcolor{red!13}~ & \cellcolor{red!100}~ & \cellcolor{red!43}~ & \cellcolor{red!37}~ & \cellcolor{red!40}~ & \cellcolor{red!19}~ &  & \cellcolor{red!100}~ & \cellcolor{red!100}~ & \cellcolor{red!5}~ & \cellcolor{red!100}~ & \cellcolor{red!43}~ & \cellcolor{red!6}~ \\
\rule{0pt}{3ex} & \multicolumn{6}{c}{\textsf{PGP}} &  & \multicolumn{6}{c}{\textsf{Gnutella}} \\
$g2_1$ & \cellcolor{red!100}~ & \cellcolor{black!25}~ & \cellcolor{red!8}~ & \cellcolor{red!92}~ & \cellcolor{red!100}~ & \cellcolor{red!4}~ &  & \cellcolor{red!100}~ & \cellcolor{red!100}~ & \cellcolor{black!25}~ & \cellcolor{red!25}~ & \cellcolor{red!100}~ & \cellcolor{red!4}~ \\
$g3_1$ & \cellcolor{red!0}~ & \cellcolor{black!25}~ & \cellcolor{red!4}~ & \cellcolor{red!31}~ & \cellcolor{red!100}~ & \cellcolor{red!4}~ &  & \cellcolor{red!35}~ & \cellcolor{red!8}~ & \cellcolor{black!25}~ & \cellcolor{red!28}~ & \cellcolor{red!100}~ & \cellcolor{red!0}~ \\
$g3_2$ & \cellcolor{red!49}~ & \cellcolor{black!25}~ & \cellcolor{red!1}~ & \cellcolor{red!7}~ & \cellcolor{red!37}~ & \cellcolor{red!18}~ &  & \cellcolor{red!100}~ & \cellcolor{red!100}~ & \cellcolor{black!25}~ & \cellcolor{red!100}~ & \cellcolor{red!44}~ & \cellcolor{red!0}~ \\
$g4_1$ & \cellcolor{red!36}~ & \cellcolor{black!25}~ & \cellcolor{red!1}~ & \cellcolor{red!7}~ & \cellcolor{red!100}~ & \cellcolor{red!6}~ &  & \cellcolor{red!11}~ & \cellcolor{red!100}~ & \cellcolor{black!25}~ & \cellcolor{red!15}~ & \cellcolor{red!100}~ & \cellcolor{red!3}~ \\
$g4_2$ & \cellcolor{red!50}~ & \cellcolor{black!25}~ & \cellcolor{red!47}~ & \cellcolor{red!2}~ & \cellcolor{red!50}~ & \cellcolor{red!22}~ &  & \cellcolor{red!100}~ & \cellcolor{red!100}~ & \cellcolor{black!25}~ & \cellcolor{red!100}~ & \cellcolor{red!50}~ & \cellcolor{red!16}~ \\
$g4_3$ & \cellcolor{red!50}~ & \cellcolor{black!25}~ & \cellcolor{red!33}~ & \cellcolor{red!77}~ & \cellcolor{red!43}~ & \cellcolor{red!17}~ &  & \cellcolor{red!100}~ & \cellcolor{red!100}~ & \cellcolor{black!25}~ & \cellcolor{red!100}~ & \cellcolor{red!49}~ & \cellcolor{red!4}~ \\
$g4_4$ & \cellcolor{red!16}~ & \cellcolor{black!25}~ & \cellcolor{red!73}~ & \cellcolor{red!100}~ & \cellcolor{red!50}~ & \cellcolor{red!29}~ &  & \cellcolor{red!100}~ & \cellcolor{red!100}~ & \cellcolor{black!25}~ & \cellcolor{red!100}~ & \cellcolor{red!9}~ & \cellcolor{red!11}~ \\
$g4_5$ & \cellcolor{red!46}~ & \cellcolor{black!25}~ & \cellcolor{red!33}~ & \cellcolor{red!36}~ & \cellcolor{red!100}~ & \cellcolor{red!18}~ &  & \cellcolor{red!100}~ & \cellcolor{red!100}~ & \cellcolor{black!25}~ & \cellcolor{red!100}~ & \cellcolor{red!100}~ & \cellcolor{red!0}~ \\
$g4_6$ & \cellcolor{red!50}~ & \cellcolor{black!25}~ & \cellcolor{red!14}~ & \cellcolor{red!3}~ & \cellcolor{red!100}~ & \cellcolor{red!11}~ &  & \cellcolor{red!82}~ & \cellcolor{red!100}~ & \cellcolor{black!25}~ & \cellcolor{red!9}~ & \cellcolor{red!100}~ & \cellcolor{red!3}~ \\
 & \textsf{~CL~} & \textsf{HRG~} & \textsf{~SBM} & \textsf{BTER} & \textsf{KRON} & \textsf{~VRG} &  & \textsf{~CL~} & \textsf{HRG~} & \textsf{~SBM} & \textsf{BTER} & \textsf{KRON} & \textsf{~VRG}
\end{tabular}
}
    }
    \caption{Relative graphlet\protect\footnotemark  ~counts as a heatmap. The color intensity in each cell indicates \textit{disagreement} between the number of graphlets found in the generated graph and the number of graphlets found in the original graph. CNRG consistently performs the best. The grayed out columns indicate that the method failed to produce graphs. This figure is best viewed in color.}
    \label{fig:heatmap}
\end{figure}

\footnotetext{Symbols are used to represent graphlet structures; $g2_1$ is an edge, $g3_1$ is a triangle, $g3_2$ is an open triangle, $g4_{1..6}$ represent clique, chordal cycle, triangle with tail, cycle, star, and path respectively~\citep{ahmed2015efficient}.}

Exponential Random Graph Models (ERGMs) are another type of graph model that learns a robust graph model from user-defined features of a graph~\citep{robins2007introduction}. Unfortunately, this model does not scale well and is prone to model degeneracy. Neural network graph models like GraphVAE~\citep{simonovsky2018graphvae} and GraphRNN~\citep{you2018graphrnn} are currently limited in their scalability. Generative adversarial networks (GANs) have been shown to scale to medium-sized graphs and perform on-par with existed methods; however, the model size of NetGAN is many times larger than the graph size~\citep{bojchevski2018netgan}. We attempted to compare these methods but were unable to because of problems with either model degeneracy or scalability.

The main purpose of node embedding models like LINE~\citep{tang2015line}, node2vec~\citep{grover2016node2vec}, VGAE~\citep{kipf2016variational}, and others~\citep{goyal2018graph} is to learn vector representations of the nodes. They are not well equipped to generate graphs and cannot be compared for this task.

\smallskip
\noindent\textbf{Evaluation.}\quad We generate 5 graphs using each model on each dataset and compare each generated graph with the original. To measure how well the local structures are preserved in the generated graph, we counted the number of size-2, 3, and 4 node graphlets~\citep{ahmed2015efficient} and compared those values to the number of graphlets present in the original graph. The (mean average) difference in graphlet counts is indicated as a heatmap in Fig.~\ref{fig:heatmap}. CNRG consistently outperforms the other models at this task.

GCD, $\lambda$-distance, and \textsc{DeltaCon} metrics are indicated in Tab.~\ref{tab:results} where bold indicates the best (mean average) performance for each dataset and metric. Across all metrics, the CNRG model performs consistently well, especially in the graphlet counts and GCD metrics. The ChungLu model does a very good job at capturing the $\lambda$-distance; this is expected because ChungLu directly (and only) models the node degree, which is highly correlated with the eigenvalues.

\section{Discussion}

The present work describes \update{CNRG, a variant of the} vertex replacement grammar model inspired by the context-free grammar formalism widely used in compilers and natural language processing. We described how a \update{CNRG} can be extracted from a hierarchical clustering of a graph and then show that the model succinctly encodes the structures present in the original graph. Starting with an empty graph, if we apply \update{CNRG} rules stochastically, then the \update{CNRG} model can generate a new graph. We show that the newly generated graphs contain global and local topographical features that are similar to the original graph. 

A potentially significant benefit from the \update{CNRG} model stems from its ability to directly encode local substructures and patterns in the RHSs of the grammar rules. Encoding these local graphlet-like structures is probably the reason that the \update{CNRG} model performed so well at the graphlet counting task and the GCD metric. Forward applications of \update{CNRG}s may allow scientists to identify previously unknown patterns in graph datasets representing important natural or physical phenomena~\citep{pennycuff2018synchronous}. Further investigation into the nature of the extracted rules and their meaning (if any) is a top priority.

We also plan to investigate differences between the grammars extracted from different types of graphs. What are the implications of finding two graphs that have a significant overlap in their extracted grammars? What about graphs that seem similar on the surface, but have little overlap in their grammar? Another area of study that we are particularly interested in is learning a temporal grammar from the dynamical processes of an evolving graph. Additional applications of \update{CNRG}s are possible on multi-level, multi-layer, and labeled graphs and their various applications.

\vspace{.2cm}
\noindent\textbf{Acknowledgements.} We thank David Chiang for his guidance on this work. This research is supported by a grant from the US National Science Foundation (\#1652492).

\small{

}
\end{document}